# Multi-shell ankylography

**Leigh S. Martin,\* Chien-Chun Chen, and Jianwei Miao**

*Department of Physics & Astronomy and California NanoSystems Institute, University of California, Los Angeles, CA 90095,USA*
*[\*Leigh.Martin@Colorado.EDU](mailto:Leigh.Martin@Colorado.EDU)*

**Abstract:** Three-dimensional (3D) imaging techniques appeal to a broad range of scientific and industrial applications. Typically, projection slice theorem enables multiple two-dimensional (2D) projections of an object to be combined in the Fourier domain to yield a 3D image. However, traditional techniques require a significant number of projections. The significant number of views required in conventional tomography not only complicates such imaging modalities, but also limits their ability to image samples that are sensitive to radiation dose or are otherwise unstable in time. In this work, we demonstrate through numerical simulations and an eigenvalue analysis that a recently developed technique called ankylography enables 3D image reconstruction using much fewer views than conventional tomography. Such a technique with the ability to obtain the 3D structure information from a few views is expected to find applications in both optical and x-ray imaging fields.

**Key words:** Coherent diffraction imaging; Ankylography; Phase retrieval.

## 1. Introduction

From tabletop high harmonic systems to free electron lasers, EUV and x-ray coherent light sources have recently undergone substantial technological leaps [1-3]. The difficulty of manufacturing high-resolution x-ray focusing optics has led to the development new imaging modalities that can take advantage of brilliant coherent x-rays made available by these advances. Coherent diffraction imaging (CDI) has proved enormously successful in imaging intricate biological, material and engineered structures [4-8]. Unlike lens-based microscopies, the spatial resolution in CDI is only limited by imaging wavelength and the diffraction angle, and is free of aberrations often introduced by focusing optics. Focal plane imaging is replaced by detection of an oversampled diffraction pattern [9], and the resulting phase problem can be solved by one of many phase retrieval algorithms [10-15].

It was demonstrated some time ago that through oversampling, phase retrieval algorithms can recover Fourier amplitudes in experimental diffraction data [4]. This was achieved by recovering both the phase information and some unmeasured data in 2D CDI, in which a beam stop or other geometric constraint prevented direct measurement of some diffraction intensity [16,17]. In 3D CDI and other tomographic reconstructions, reciprocal space points that do not intersect with the Ewald spheres have also been recovered computationally, which enabled volumetric reconstruction in the presence of a missing wedge and small gaps between data sets at high spatial frequencies [18-26].

In 2010, Raines *et al.* demonstrated a technique, termed ankylography, in which measurement of a single diffraction pattern on a spherical shell could yield full 3D reconstructions for small objects or large objects with low resolutions [27]. Computationally, ankylography is similar to tomographic CDI, but as only a single spherical shell is measured, it requires computational recovery of a much larger portion of the data. The method showed that a single oversampled diffraction pattern contains 3D information, which may be accessed through phase retrieval methods. It also demonstrated the extent to which missing Fourier amplitudes may be recovered. While a finite object's structure is in principle contained in the Ewald sphere, reconstructing objects that are large with respect to the resolution remains a significant challenge, and considerations of bandlimited information transfer prevent ankylography from generalizing to arbitrarily large objects in the presence of noise [28,29]. Numerous solutions have been proposed, including the use of broader imaging bandwidths coupled with energy resolved detectors, reference objects and *a priori* information [30-32].

## 2. Methods

In this work, we demonstrate that ankylography has the capability to be used with multiple views, much like conventional tomography. Although this application takes ankylography out of the single shot regime, insights from single view ankylography offer the potential to produce 3D images in fewer views than conventional tomography. We show that oversampled high numerical aperture diffraction patterns convey enough information to reconstruct objects in fewer views than have previously been achieved in any experimental demonstration of tomography. Furthermore, we include an eigenvalue analysis which demonstrates that the improvement offered by multi-shell ankylography generalizes beyond the two examples presented here. While our numerical simulations were performed in the optical regime, they are equally applicable to any coherent system and object in which the Born approximation is satisfied.

Multi-shell ankylography and tomography record disparate types of information about the target object. Conventional tomography records a flat, paraxial portion of the Ewald sphere, and consequently a single view only provides information in directions perpendicular to the viewing angle. In other words, acquired data depend only on the integral of the object along the viewing axis and information along this third dimension is unavailable. In contrast, ankylographic diffraction patterns derive from the complete 3D structure of the imaged sample and no uncoupling of dimensions can be made. Figure 1 shows one diffraction pattern for each imaging modality. The ankylographic image not only shows distortion due to curvature of the Ewald sphere, but also shows different features arising from the sampling of different regions of reciprocal space.

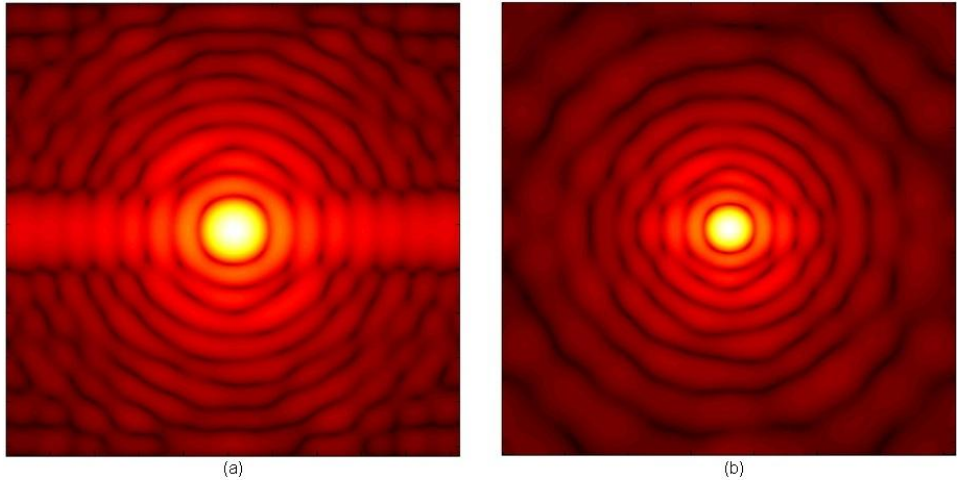

Fig. 1. (a) A diffraction pattern of the scaled, clipped Shepp Logan phantom where the diffraction angle is very small and the curvature of Ewald sphere is negligible. (b) An ankylographic diffraction pattern of the same object as it would appear on a flat detector where the curvature of Ewald sphere is not negligible. Intensity values have been raised to the 1/8 power for visualization.

As a proof-of-concept demonstration, we performed reconstructions using the two diffraction patterns shown in Fig. 1. These correspond to imaging a 20x20x20 voxel scaled Shepp Logan phantom shown in Fig. 2. A small portion of the phantom was removed to break inversion symmetry of the sample's support. Inversion symmetric objects have been shown to be more difficult to reconstruct in CDI [9], and in this case no good reconstructions were obtained with a symmetric object. Figure 2 also shows reconstructions of the phantoms for tomography, ankylography and an intermediate case using two views each. Physically, these reconstructions correspond to imaging the same object while scaling the numerical aperture and wavelength so as to maintain constant resolution. As the oversampling ratio was chosen to maintain an equal number of known voxels in all cases [9,27], the dramatic improvement in reconstruction quality in multi-shell ankylography is due to the curvature of the Ewald sphere.

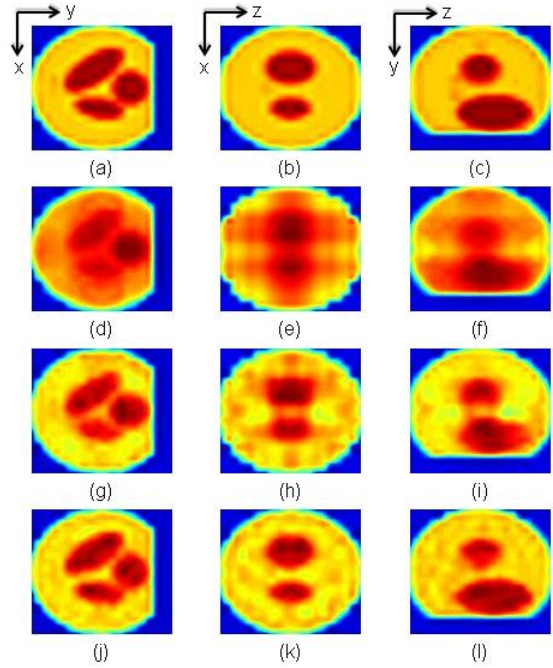

Fig. 2. (a)-(c) XY, XZ and YZ slices of the original Shepp Logan phantom sample. A small portion of the phantom was removed to break inversion symmetry of the sample's support. Inversion symmetric objects are generally difficult to reconstruct in CDI [9]. (d)-(f) Reconstruction of the corresponding slices using only an approximately flat subset of the Ewald sphere. (g)-(i) Reconstruction with a maximum diffraction angle of 12°. (j)-(l) Reconstruction using the entire forward scattering portion of the Ewald sphere. In all cases, two views were used to reconstruct the object: one along the x direction and one along the z direction.

Next, we simulated multi-shell ankylography using a model pollen grain comprised of 32x32x32 voxels embedded in a 128x128x128 voxel array. Results correspond to an optical laser illumination with λ=543 nm. To simulate imaging, we Fourier transformed the 3D array, eliminated the phase information and then took all points off the Ewald spheres to be unknown values. We also eliminated data from a 1.5° missing center to account for the presence of a beam stop. The Ewald spheres consist of one voxel thick spherical shells extending to a diffraction angle $2\theta$ of 39° at the boundary. For optimal reconstruction quality, we allowed the shells to extend to the edge of the reciprocal space grid, as opposed to limiting all scattering angles to 39° or less. Experimentally, this necessitates the measurements of diffraction patterns that extend up to 63° degrees, depending on the orientation of the shell in the reciprocal space array. Lastly, we added Poisson noise corresponding to a total detected flux of $\sim 10^{12}$ photons per diffraction pattern. Another set of reconstructions used only $\sim 10^9$ photons to investigate the effect of a lower signal-to-noise ratio (SNR). The simulation does not incorporate interpolation errors between the measured diffraction pattern and the computation grid. Although compensation for these errors is still an active area of research in ankylography, such problems have already been ameliorated experimentally by using a higher oversampling ratio [27,30], which reduces the spacing between the Ewald sphere and sampled points in discrete reciprocal space. As the object is finitely supported, the resulting diffraction patterns are band limited and thus the already small interpolation errors diminish as the oversampling degree increases. It should also be noted that these errors were present and well tolerated in a range of ankylography and coherent diffraction tomography experiments [18-

24,27,30], and that similar errors result from the finite bandwidth of lasers used in all experimental CDI.

As samples must often be mounted on a membrane, we simulated equally spaced projections between −45° and 45°. As we treat the sample as a real object, the Fourier transform is centro-symmetric, and consequently each view contributes two shells inverted through the origin. To quantify the number of shells required to image our object, we ran twenty-four independent reconstructions of the object for each case, from two to six shells. Each reconstruction is seeded by running 100 different initial phase guesses and selecting the best one for the final computation. We rank the quality of the seeds based on the difference between the measured and known diffraction patterns. The best seed was then iterated 5000 times with the 3D Hybrid Input Output (HIO) algorithm to obtain a final reconstruction [10]. Although the object contains non-scattering regions within its bounding box, we used a cubic support. This looser support represents a worst case scenario for an object with the given size relative to the imaging wavelength. In practice, the use of algorithms such as shrink-wrap can actively provide a better estimate of the support, offering improved reconstructions [33]. Furthermore, an effective iterative algorithm, termed Oversampling Smoothness (OSS), has recently been developed [15]. Both numerical simulations and experimental results indicate that OSS outperforms HIO in terms of accuracy and consistency for phase retrieval. The application of OSS to multi-shell ankylographic data will be explored in a follow-up study.

Reconstruction quality can be judged in numerous ways and different applications favor different methods. In experiment, measures must be based solely on available data, and are typically based on comparison with the original diffraction data. However, in sparse recovery, such error metrics do not necessarily correlate with reconstruction quality due to the potential existence of weakly constrained modes, or modes that may be added to the reconstruction without dramatically changing consistency with measured data. Simulation offers the unique advantage that the final image may be directly compared with the original, enabling direct measure of the presence of such artifacts. Numerical simulation also circumvents issues of comparison between imaging modalities with different transfer functions, which can be subtle and highly subjective. Thus in addition to a reciprocal space based error metric, we quantify reconstruction quality by calculating the normalized RMS difference from a central patch of the object (26 × 26 × 26 voxels). Using this subset allows calculation of the error when the reconstruction is shifted within the support.

Simulation alone does not justify the claim that multi-shell ankylography always surpasses conventional tomography. To further support this claim, we develop an eigenvalue analysis related to that employed in refs. [17,34]. Such analyses address the issue of uniqueness by searching for minimally constrained modes that have small or zero amplitude in the set of known voxels in reciprocal space but non-zero values elsewhere. Thus these modes can be added to a reconstruction while maintaining consistency with recorded diffraction data.

We find these modes using the following operators, which act on a real-space voxel array $m$. Let $P_S$ be an operator which sets all voxels outside a given support S to zero. Let $\tilde{P}_E$ be an operator which Fourier transforms $m$, sets its amplitudes to zero outside the set $E$ of voxels on the Ewald spheres and then inverse Fourier transforms. As we will be interested in diagonalizing a matrix, we wish to work with as few voxels as possible to reduce the complexity of the calculation. Choosing the operator $P_S \tilde{P}_E P_S$ allows us to work with only the subspace $S$ of the full voxel array as its elements are zero outside of $S$. Furthermore, $P_S \tilde{P}_E P_S$ has the relevant property that for all m supported on $S$:

$$\langle m | P_S \tilde{P}_E P_S | m \rangle = \langle P_S[m] | \tilde{P}_E | P_S[m] \rangle \tag{1}$$

$$= \langle m | \mathrm{F}^{-1} P_E \mathrm{F} | m \rangle \tag{2}$$

$$= \langle \mathrm{F} m | P_E | \mathrm{F} m \rangle \tag{3}$$

$$= \sum_{\mathbf{k} \in E} |\mathrm{F}\,[m](\mathbf{k})|^2 \qquad (4)$$

where $\mathrm{F}$ is the discrete Fourier transform. This result shows that the eigenvalues of $M = P_S \tilde{P}_E P_S$ give the fraction of amplitudes of $\mathrm{F}\,[m]$ that lie on the Ewald spheres. It is simple to see that $M$ is Hermitian and thus a complete orthonormal eigenvector basis can be found. Using this orthogonally it can be shown that the least constrained mode corresponds to the eigenvector of $M$ with the smallest eigenvalue. Thus the eigenvalues of $M$ give the constraining power of the corresponding mode, and small eigenvectors indicate greater ambiguity in the best possible reconstruction in the presence of noise.

### 3. Results

Reconstruction of the pollen grain using 2, 3, 4, 5 and 6 views all show the salient features of the pollen grain sample. As shown in Fig. 3, the object is clear and free of noticeable artifacts when four shells are used. Fourier domain phases, amplitudes between shells and amplitudes in the missing cone were retrieved *ab initio*. Note that all data could be imaged using a conventional flat detector. Figure 4 shows the real space error as a function of the number of shells, which decreases dramatically with each additional view. The sample is discernible with only two shells (not shown here), a result that would not be possible in conventional tomography. The reconstruction achieves excellent quality for only four shells. Resolution in all three dimensions is 0.86 µm, as determined by the diffraction angle and imaging wavelength.

Shown in the last row of Fig. 3 is the same sample reconstructed with greater noise in the diffraction pattern. The appearance of the reconstructions does not change noticeably, with only minor, tolerable changes appearing. The real-space error shows the same trend visible in Fig. 3. The change in quality indicates the presence of noise in the reconstruction correlates with the noise level in the diffraction data, showing no signs of fundamental instability arising as a result of a reduced SNR.

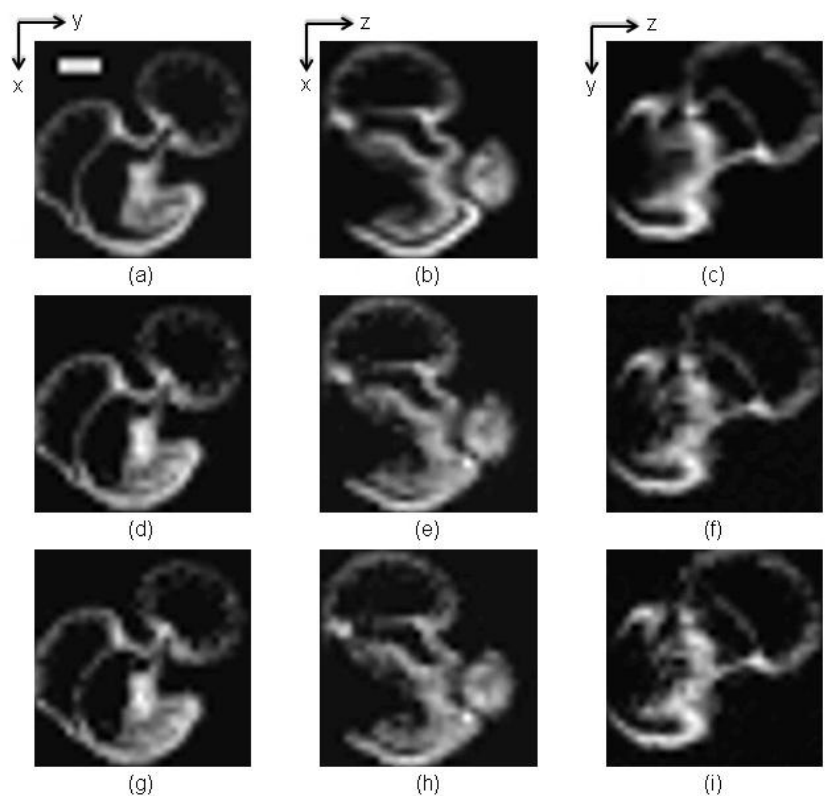

Fig. 3. (a)-(c) XY, XZ and YZ views of the original pollen grain sample. Scale bar is 5 μm. (d)-(f) Corresponding views of the best reconstruction in the presence of low noise, with a detected flux of $10^{12}$ photons per diffraction pattern using four shells. (g)-(h) Corresponding views of the best reconstruction in the presence of high noise, with a detected flux of $10^9$ photons per diffraction pattern using four shells. The fine features in the reconstructed pollen grain are in good agreement with those in the model.

The physical structure of a sample determines reconstruction quality. Even two objects with an identical number of unknown voxels will not necessarily reconstruct the same. Consideration of weakly constrained modes gives some insight into this phenomenon. For example, in order to maintain consistency with the measured DC component (when measured), such modes must contain negative values. If an object contains zero regions where a weakly constrained mode is negative, then the non-negativity constraint in ankylography can rule out contribution of this mode. Thus even with a loose support, sparsity should play an important role in multi-shell ankylography. It should also be noted that as data sets are sparser at higher spatial frequencies, objects with sharp transitions in scattering amplitude may prove more difficult to reconstruct.

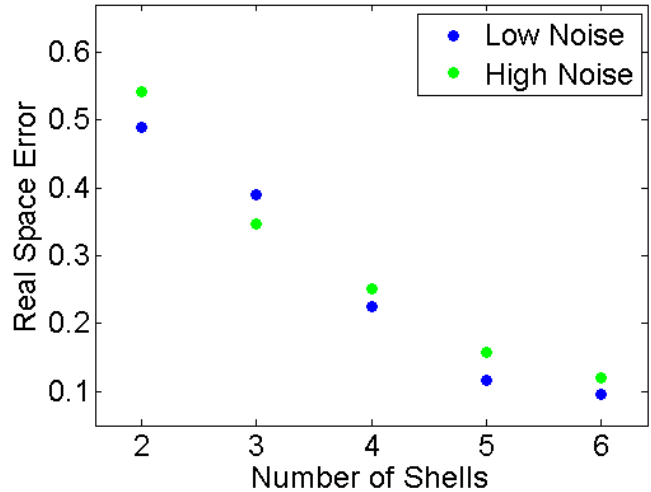

Fig. 4. Mean real space error as a function of the number of shells used in the reconstructions. The real space error derives from the RMS difference between the model and the best reconstruction, where the best reconstruction is determined based on reciprocal space RMS error. Low noise corresponds to $10^{12}$ photons per diffraction pattern and high noise corresponds to $10^9$ photons.

Our eigenvalue analysis varies numerical aperture, projection number and object size over a range of values to explore a larger parameter set than that which is covered in the simulations. As matrix size scales with the number of voxels inside the support, we limit our calculations to objects that are 17x17x17 voxels or smaller. With more computational resources, the eigenvalue analysis can, in principle, be extended to larger objects. As derived above, eigenvalues give an indication of how much uncertainty in measured data gets amplified to contribute to uncertainty in reconstructions; smaller eigenvalues correspond to greater ambiguity.

Figure 5 shows the eigenvalue spectrum resulting from 6 shells used to image an 11x11x11 voxel object. The figure shows the eigenvalues as a function of maximum diffraction angle and two 2D plots for quantitative interpretation of the 3D image. The case of $2\theta=0$ gives by far the lowest eigenvalues. We observe this trend consistently in all 18 cases calculated, which include 2, 4, 6 and 7 shells as well as a variety of object sizes and oversampling ratios. In some cases, the eigenvalues for small $2\theta$ values were negative. Equation 4 shows that the eigenvalues must be positive, and thus we attribute negative eigenvalues to numerical errors arising from finite machine precision. This situation arises when the eigenvalues become very close to zero. Another interesting feature that reproduces in all data sets concerning more than 2 projections is the local maximum at around 10°. This result shows that the benefits of multi-shell ankylography may be obtained even at low numerical aperture, potentially simplifying such experiments. This maximum is sometimes global but is typically not.

The presence of small eigenvalues shows that even for large numbers of projections, one should expect the SNR to be much smaller in the object domain than in the Fourier domain. However as phase retrieval generally requires a high Fourier domain SNR to achieve a moderate object domain SNR, this result alone is not indicative of a problem. The eigenvalue spectra corresponding to our simulations contain even smaller eigenvalues and nevertheless do not preclude excellent reconstructions. Nevertheless, the fundamental limit of ankylography and related techniques is strongly linked to the rapid decrease in eigenvalue magnitude with the increase of the object size. This analysis is intended as an analytic step toward definitively determining the set of conditions under which ankylography will or will not work. Furthermore, knowing the least constrained modes could prove useful in the reconstruction process and in incorporating *a priori* information into recovered objects.

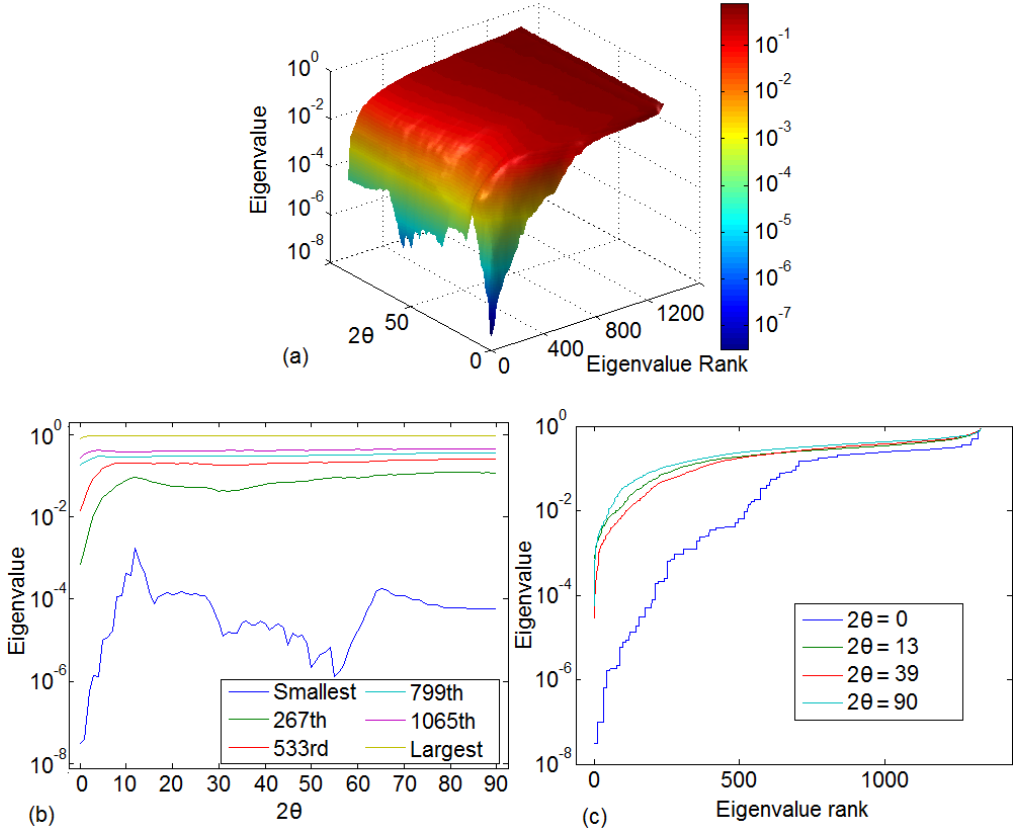

Fig. 5. (a) Eigenvalues for the case of an 11x11x11 voxel object embedded in a 48x48x48 voxel grid containing 6 equally spaced shells. Results are plotted on a logarithmic scale. Eigenvalues are ranked from 1 to 1331 by their magnitude. (b) The smallest eigenvalue, largest and four intermediate eigenvalues plotted as a function of $2\theta$. (c) Full eigenvalue spectra for four values of $2\theta$. The jagged structure observed when $2\theta=0$ is a consistently observed feature. Note that the eigenvalues for multi-shell ankylography are up to 4 orders of magnitude larger those for paraxial tomography.

## 4. Conclusion

In conventional tomography, each projection is flat, and therefore any one projection contains no information about structure in the direction of the projection. In contrast, each ankylographic shell contains 3D information about the object, and thus constrains reciprocal space amplitudes throughout the entire volume. One practical implication of our results is that if one scales the numerical aperture and imaging wavelength so as to maintain constant imaging resolution, increasing Ewald sphere curvature allows a reduction in the number of views or better reconstruction. More generally, we note the importance of Ewald sphere curvature in constraining the reconstruction using known data and show that when working in this regime, the number of view can be reduced. As tomographic CDI is sometimes limited by sample degradation, these results may extend the types of objects that can be three-dimensionally imaged with tabletop HHG sources, EUV lasers, 3rd generation synchrotron radiation and other coherent x-ray sources, provided that high numerical aperture data can be collected in a feasible geometry. Furthermore, the reduction in the number of required views enables more simple and efficient data acquisition.

Our eigenvalue analysis not only confirms the advantages of multi-shell ankylography but also provide a useful step in rigorously deriving the fundamental limitations and capabilities of all forms of ankylography and tomographic CDI. The above simulations demonstrate the utility of multi-shell ankylography for any useful imaging wavelength. Our results show that biological samples like the one reconstructed above could be three-dimensionally imaged at a resolution comparable to that available in confocal microscopy without the use of stains or fluorescent tags. For imaging at shorter wavelengths, multi-shell ankylography may provide a tool to obtain more information from a sample before it is degraded by the ionizing radiation used. In addition, this work opens avenues for a new kind of multiple-wavelength tomography, as in ankylography each wavelength in principle represents another channel of communication. 3D biological imaging has also recently been implemented in inexpensive devices with applications outside the lab [35], and multi-shell ankylography may have applications in improving their simplicity, robustness and portability. By reconstructing a 3D sample from very few projections, it is hoped that this ability will prove useful in a broad range of imaging fields.

**Acknowledgments**

We thank Margaret M. Murnane and Henry C. Kapteyn for stimulating discussions. This work was conducted at the REU UCLA Physics & Astrophysics Summer Program funded by National Science Foundation.